\PassOptionsToPackage{unicode}{hyperref}
\PassOptionsToPackage{hyphens}{url}
\PassOptionsToPackage{dvipsnames,svgnames,x11names}{xcolor}
\documentclass[
]{article}
  \usepackage[utf8]{inputenc}
  \usepackage{newunicodechar}
  \newunicodechar{σ}{\ensuremath{\sigma}}
\usepackage{amsmath,amssymb}
\usepackage{lmodern}
\usepackage{iftex}
\ifPDFTeX
  \usepackage[T1]{fontenc}
  \usepackage[utf8]{inputenc}
  \usepackage{textcomp} 
\else 
  \usepackage{unicode-math}
  \defaultfontfeatures{Scale=MatchLowercase}
  \defaultfontfeatures[\rmfamily]{Ligatures=TeX,Scale=1}
\fi
\IfFileExists{upquote.sty}{\usepackage{upquote}}{}
\IfFileExists{microtype.sty}{
  \usepackage[]{microtype}
  \UseMicrotypeSet[protrusion]{basicmath} 
}{}
\makeatletter
\@ifundefined{KOMAClassName}{
  \IfFileExists{parskip.sty}{%
    \usepackage{parskip}
  }{
    \setlength{\parindent}{0pt}
    \setlength{\parskip}{6pt plus 2pt minus 1pt}}
}{
  \KOMAoptions{parskip=half}}
\makeatother
\usepackage{xcolor}
\usepackage{color}
\usepackage{fancyvrb}

\DefineVerbatimEnvironment{Highlighting}{Verbatim}{commandchars=\\\{\}}
\newenvironment{Shaded}{}{}

\newcommand{\CommentTok}[1]{\textcolor[rgb]{0.38,0.63,0.69}{\textit{#1}}}

\newcommand{\DecValTok}[1]{\textcolor[rgb]{0.25,0.63,0.44}{#1}}

\newcommand{\NormalTok}[1]{#1}
\newcommand{\OperatorTok}[1]{\textcolor[rgb]{0.40,0.40,0.40}{#1}}

\newcommand{\StringTok}[1]{\textcolor[rgb]{0.25,0.44,0.63}{#1}}

\usepackage{longtable,booktabs,array}
\usepackage{calc} 
\usepackage{etoolbox}
\makeatletter
\patchcmd\longtable{\par}{\if@noskipsec\mbox{}\fi\par}{}{}
\makeatother
\IfFileExists{footnotehyper.sty}{\usepackage{footnotehyper}}{\usepackage{footnote}}
\makesavenoteenv{longtable}
\providecommand{\tightlist}{%
  \setlength{\itemsep}{0pt}\setlength{\parskip}{0pt}}
\NewDocumentCommand\citeproctext{}{}
\NewDocumentCommand\citeproc{mm}{%
  \begingroup\def\citeproctext{#2}\cite{#1}\endgroup}
\makeatletter
 \let\@cite@ofmt\@firstofone
 \def\@biblabel#1{}
 \def\@cite#1#2{{#1\if@tempswa , #2\fi}}
\makeatother
\newlength{\cslhangindent}
\newlength{\csllabelwidth}
\newenvironment{CSLReferences}[2] 
 {\begin{list}{}{%
  \setlength{\itemindent}{0pt}
  \setlength{\leftmargin}{0pt}
  \setlength{\parsep}{0pt}
  \ifodd #1
   \setlength{\leftmargin}{\cslhangindent}
   \setlength{\itemindent}{-1\cslhangindent}
  \fi
  \setlength{\itemsep}{#2\baselineskip}}}
 {\end{list}}
\usepackage{calc}

\ifLuaTeX
\usepackage[bidi=basic]{babel}
\else
\usepackage[bidi=default]{babel}
\fi
\babelprovide[main,import]{american}

\def\languageshorthands#1{}
\ifLuaTeX
  \usepackage{selnolig}  
\fi
\IfFileExists{bookmark.sty}{\usepackage{bookmark}}{\usepackage{hyperref}}
\IfFileExists{xurl.sty}{\usepackage{xurl}}{} 
\hypersetup{
  pdftitle={Furax: A Modular JAX Framework for Linear Operators in
Astrophysical and Cosmological Data Analysis},
  pdfauthor={Pierre Chanial, Simon Biquard, Wassim Kabalan, Wuhyun Sohn,
Artem Basyrov, Benjamin Beringue, Alexandre Boucaud, Andréa Landais,
Magdy Morshed, Radek Stompor, Ema Tsang King Sang, Amalia
Villarrubia-Aguilar, Josquin Errard},
  pdflang={en-US},
  colorlinks=true,
  linkcolor={Maroon},
  filecolor={Maroon},
  citecolor={Blue},
  urlcolor={Blue},
  pdfcreator={LaTeX via pandoc}}

\title{Furax: A Modular JAX Framework for Linear Operators in
Astrophysical and Cosmological Data Analysis}

\definecolor{c53baa1}{RGB}{83,186,161}
\definecolor{c202826}{RGB}{32,40,38}
\def \rorglobalscale {0.1}
\newcommand{\rorlogo}{%
\begin{tikzpicture}[y=1cm, x=1cm, yscale=\rorglobalscale,xscale=\rorglobalscale, every node/.append style={scale=\rorglobalscale}, inner sep=0pt, outer sep=0pt]
  \begin{scope}[even odd rule,line join=round,miter limit=2.0,shift={(-0.025, 0.0216)}]
    \path[fill=c53baa1,nonzero rule,line join=round,miter limit=2.0] (1.8164, 3.012) -- (1.4954, 2.5204) -- (1.1742, 3.012) -- (1.8164, 3.012) -- cycle;
    \path[fill=c53baa1,nonzero rule,line join=round,miter limit=2.0] (3.1594, 3.012) -- (2.8385, 2.5204) -- (2.5172, 3.012) -- (3.1594, 3.012) -- cycle;
    \path[fill=c53baa1,nonzero rule,line join=round,miter limit=2.0] (1.1742, 0.0669) -- (1.4954, 0.5588) -- (1.8164, 0.0669) -- (1.1742, 0.0669) -- cycle;
    \path[fill=c53baa1,nonzero rule,line join=round,miter limit=2.0] (2.5172, 0.0669) -- (2.8385, 0.5588) -- (3.1594, 0.0669) -- (2.5172, 0.0669) -- cycle;
    \path[fill=c202826,nonzero rule,line join=round,miter limit=2.0] (3.8505, 1.4364).. controls (3.9643, 1.4576) and (4.0508, 1.5081) .. (4.1098, 1.5878).. controls (4.169, 1.6674) and (4.1984, 1.7642) .. (4.1984, 1.8777).. controls (4.1984, 1.9719) and (4.182, 2.0503) .. (4.1495, 2.1132).. controls (4.1169, 2.1762) and (4.0727, 2.2262) .. (4.0174, 2.2635).. controls (3.9621, 2.3006) and (3.8976, 2.3273) .. (3.824, 2.3432).. controls (3.7505, 2.359) and (3.6727, 2.367) .. (3.5909, 2.367) -- (2.9676, 2.367) -- (2.9676, 1.8688).. controls (2.9625, 1.8833) and (2.9572, 1.8976) .. (2.9514, 1.9119).. controls (2.9083, 2.0164) and (2.848, 2.1056) .. (2.7705, 2.1791).. controls (2.6929, 2.2527) and (2.6014, 2.3093) .. (2.495, 2.3487).. controls (2.3889, 2.3881) and (2.2728, 2.408) .. (2.1468, 2.408).. controls (2.0209, 2.408) and (1.905, 2.3881) .. (1.7986, 2.3487).. controls (1.6925, 2.3093) and (1.6007, 2.2527) .. (1.5232, 2.1791).. controls (1.4539, 2.1132) and (1.3983, 2.0346) .. (1.3565, 1.9436).. controls (1.3504, 2.009) and (1.3351, 2.0656) .. (1.3105, 2.1132).. controls (1.2779, 2.1762) and (1.2338, 2.2262) .. (1.1785, 2.2635).. controls (1.1232, 2.3006) and (1.0586, 2.3273) .. (0.985, 2.3432).. controls (0.9115, 2.359) and (0.8337, 2.367) .. (0.7519, 2.367) -- (0.1289, 2.367) -- (0.1289, 0.7562) -- (0.4837, 0.7562) -- (0.4837, 1.4002) -- (0.6588, 1.4002) -- (0.9956, 0.7562) -- (1.4211, 0.7562) -- (1.0118, 1.4364).. controls (1.1255, 1.4576) and (1.2121, 1.5081) .. (1.2711, 1.5878).. controls (1.2737, 1.5915) and (1.2761, 1.5954) .. (1.2787, 1.5991).. controls (1.2782, 1.5867) and (1.2779, 1.5743) .. (1.2779, 1.5616).. controls (1.2779, 1.4327) and (1.2996, 1.3158) .. (1.3428, 1.2113).. controls (1.3859, 1.1068) and (1.4462, 1.0176) .. (1.5237, 0.944).. controls (1.601, 0.8705) and (1.6928, 0.8139) .. (1.7992, 0.7744).. controls (1.9053, 0.735) and (2.0214, 0.7152) .. (2.1474, 0.7152).. controls (2.2733, 0.7152) and (2.3892, 0.735) .. (2.4956, 0.7744).. controls (2.6016, 0.8139) and (2.6935, 0.8705) .. (2.771, 0.944).. controls (2.8482, 1.0176) and (2.9086, 1.1068) .. (2.952, 1.2113).. controls (2.9578, 1.2253) and (2.9631, 1.2398) .. (2.9681, 1.2544) -- (2.9681, 0.7562) -- (3.3229, 0.7562) -- (3.3229, 1.4002) -- (3.4981, 1.4002) -- (3.8349, 0.7562) -- (4.2603, 0.7562) -- (3.8505, 1.4364) -- cycle(0.9628, 1.7777).. controls (0.9438, 1.7534) and (0.92, 1.7357) .. (0.8911, 1.7243).. controls (0.8623, 1.7129) and (0.83, 1.706) .. (0.7945, 1.7039).. controls (0.7588, 1.7015) and (0.7252, 1.7005) .. (0.6932, 1.7005) -- (0.4839, 1.7005) -- (0.4839, 2.0667) -- (0.716, 2.0667).. controls (0.7477, 2.0667) and (0.7805, 2.0643) .. (0.8139, 2.0598).. controls (0.8472, 2.0553) and (0.8768, 2.0466) .. (0.9025, 2.0336).. controls (0.9282, 2.0206) and (0.9496, 2.0021) .. (0.9663, 1.9778).. controls (0.9829, 1.9534) and (0.9914, 1.9209) .. (0.9914, 1.8799).. controls (0.9914, 1.8362) and (0.9819, 1.8021) .. (0.9628, 1.7777) -- cycle(2.6125, 1.3533).. controls (2.5889, 1.2904) and (2.5553, 1.2359) .. (2.5112, 1.1896).. controls (2.4672, 1.1433) and (2.4146, 1.1073) .. (2.3529, 1.0814).. controls (2.2916, 1.0554) and (2.2228, 1.0427) .. (2.1471, 1.0427).. controls (2.0712, 1.0427) and (2.0026, 1.0557) .. (1.9412, 1.0814).. controls (1.8799, 1.107) and (1.8272, 1.1433) .. (1.783, 1.1896).. controls (1.7391, 1.2359) and (1.7052, 1.2904) .. (1.6817, 1.3533).. controls (1.6581, 1.4163) and (1.6465, 1.4856) .. (1.6465, 1.5616).. controls (1.6465, 1.6359) and (1.6581, 1.705) .. (1.6817, 1.7687).. controls (1.7052, 1.8325) and (1.7388, 1.8873) .. (1.783, 1.9336).. controls (1.8269, 1.9799) and (1.8796, 2.0159) .. (1.9412, 2.0418).. controls (2.0026, 2.0675) and (2.0712, 2.0804) .. (2.1471, 2.0804).. controls (2.223, 2.0804) and (2.2916, 2.0675) .. (2.3529, 2.0418).. controls (2.4143, 2.0161) and (2.467, 1.9799) .. (2.5112, 1.9336).. controls (2.5551, 1.8873) and (2.5889, 1.8322) .. (2.6125, 1.7687).. controls (2.636, 1.705) and (2.6477, 1.6359) .. (2.6477, 1.5616).. controls (2.6477, 1.4856) and (2.636, 1.4163) .. (2.6125, 1.3533) -- cycle(3.8015, 1.7777).. controls (3.7825, 1.7534) and (3.7587, 1.7357) .. (3.7298, 1.7243).. controls (3.701, 1.7129) and (3.6687, 1.706) .. (3.6333, 1.7039).. controls (3.5975, 1.7015) and (3.5639, 1.7005) .. (3.5319, 1.7005) -- (3.3226, 1.7005) -- (3.3226, 2.0667) -- (3.5547, 2.0667).. controls (3.5864, 2.0667) and (3.6192, 2.0643) .. (3.6526, 2.0598).. controls (3.6859, 2.0553) and (3.7155, 2.0466) .. (3.7412, 2.0336).. controls (3.7669, 2.0206) and (3.7883, 2.0021) .. (3.805, 1.9778).. controls (3.8216, 1.9534) and (3.8301, 1.9209) .. (3.8301, 1.8799).. controls (3.8301, 1.8362) and (3.8206, 1.8021) .. (3.8015, 1.7777) -- cycle;
  \end{scope}
\end{tikzpicture}
}

\usepackage[affil-it]{authblk}
\usepackage{orcidlink}
\author[1%
  \ensuremath\mathparagraph]{Pierre Chanial%
    \,\orcidlink{0000-0003-1753-524X}\,%
    }
\author[2%
  ]{Simon Biquard%
    \,\orcidlink{0000-0002-1493-2963}\,%
    }
\author[1%
  ]{Wassim Kabalan%
    \,\orcidlink{0009-0001-6501-4564}\,%
    }
\author[1%
  ]{Wuhyun Sohn%
    \,\orcidlink{0000-0002-6039-8247}\,%
    }
\author[1%
  ]{Artem Basyrov%
    \,\orcidlink{0000-0002-4365-4405}\,%
    }
\author[1%
  ]{Benjamin Beringue%
    \,\orcidlink{0000-0001-9571-6148}\,%
    }
\author[1%
  ]{Alexandre Boucaud%
    \,\orcidlink{0000-0001-7387-2633}\,%
    }
\author[1%
  ]{Andréa Landais%
    \,\orcidlink{0009-0005-8436-1333}\,%
    }
\author[3%
  ]{Magdy Morshed%
    \,\orcidlink{0000-0002-3214-8881}\,%
    }
\author[1%
  ]{Radek Stompor%
    \,\orcidlink{0000-0002-9777-3813}\,%
    }
\author[1%
  ]{Ema Tsang King Sang%
    \,\orcidlink{0009-0001-6108-9518}\,%
    }
\author[1%
  ]{Amalia Villarrubia-Aguilar%
    \,\orcidlink{0009-0004-4775-9935}\,%
    }
\author[1%
  \ensuremath\mathparagraph]{Josquin Errard%
    \,\orcidlink{0000-0002-1419-0031}\,%
    }

\affil[1]{Université Paris Cité, CNRS, Astroparticule et Cosmologie,
F-75013 Paris, France%
    \,\protect\href{https://ror.org/03tnjrr49}{\protect\rorlogo}\,%
  }
\affil[2]{Jodrell Bank Centre for Astrophysics, The University of
Manchester, Oxford Road, Manchester M13 9PL, UK%
    \,\protect\href{https://ror.org/027m9bs27}{\protect\rorlogo}\,%
  }
\affil[3]{Istituto Nazionale di Fisica Nucleare, Sezione di Ferrara, via
Saragat 1, I-44122 Ferrara, Italy%
    \,\protect\href{https://ror.org/00zs3y046}{\protect\rorlogo}\,%
  }
\affil[$\mathparagraph$]{Corresponding author: %
}
\date{21 March 2026}

\begin{document}
\maketitle

\section{Summary}\label{summary}

The \emph{Framework for Unified and Robust data Analysis with JAX}
(\texttt{Furax}) is an open-source Python framework for modeling data
acquisition systems and solving inverse problems in astrophysics and
cosmology. Built on \texttt{JAX} (\citeproc{ref-jax2018}{Bradbury et
al., 2018}), \texttt{Furax} provides composable building blocks in the
form of general-purpose and domain-specific linear operators, along with
pre-conditioners and solvers for their numerical inversion.
Domain-specific tools are provided for astrophysical and cosmic
microwave background (CMB) data analysis---including map-making,
instrument modeling, and astrophysical component separation---with a
modular architecture designed to extend to other fields.

\texttt{Furax} fully utilises \texttt{JAX}'s just-in-time compilation
and automatic differentiation to achieve competitive performance,
further accelerated using GPUs or TPUs. With \texttt{Furax}, researchers
can rapidly prototype and validate analysis pipelines with
production-ready computational efficiency.

\texttt{Furax} is hosted on
\href{https://github.com/CMBSciPol/furax}{GitHub}, installable via
\href{https://pypi.org/project/furax}{PyPI} and documented on
\href{https://furax.readthedocs.io}{Read the Docs}.

\section{Statement of Need}\label{statement-of-need}

Contemporary and future CMB experiments such as the Simons Observatory
(\citeproc{ref-simons2019}{The Simons Observatory Collaboration, 2019}),
the South Pole Observatory (\citeproc{ref-spo}{KIPAC, 2026}), QUBIC
(\citeproc{ref-qubic2022}{collaboration et al., 2022}) and
\emph{LiteBIRD} (\citeproc{ref-litebird2023}{LiteBIRD Collaboration et
al., 2023}) will generate massive time-ordered data (TOD) streams that
must be processed to extract cosmological information. A central problem
in CMB data analysis is to exploit data acquisition redundancy through
map-making, i.e.~recovering the sky signal \(\mathbf{m}\) (and
potentially separating its components) from noisy observations
\(\mathbf{d}\) through the linear model

\[\mathbf{d} = \mathbf{H}\mathbf{m} + \mathbf{T}\mathbf{t} + \mathbf{n}\]

where \(\mathbf{H}\) represents the data acquisition system, encoding
the pointing matrix, instrument response, and other effects;
\(\mathbf{T}\) denotes the unwanted additive effects, which need to be
removed in the analysis process, and n is the noise. A general solution
can be then obtained in a form of a generalized least-squares estimator,
(\citeproc{ref-poletti2017}{Poletti et al., 2017}):

\[\hat{\mathbf{m}} = (\mathbf{H}^\top \mathbf{F_T}^{-1} \mathbf{H})^{-1} \mathbf{H}^\top \mathbf{F_T}^{-1} \mathbf{d}\]

where \(\mathbf{F_T}\) is a weighting and deprojection operator defined
as,

\[\mathbf{F_T} = \mathbf{W} - \mathbf{W} \mathbf{T} (\mathbf{T}^\top \mathbf{W} \mathbf{T})^{-1} \mathbf{T}^\top\mathbf{W}\]

for some suitable chosen, positively defined weights, \(\mathbf{W}\).
All such solutions require efficient application of the acquisition
operator and its transpose, and would benefit from a framework
supporting operator algebra.

Historically, many data reduction pipelines developed by large
collaborations have been tied to specific experiments and did not
outlive them, often due to the lack of generality, reliance on legacy
technologies or evolving hardware paradigms. \texttt{Furax} aims to
break this pattern by being experiment-agnostic and built on Python and
\texttt{JAX}---a modern, sustainable foundation.

\texttt{Furax} addresses the above challenges by:

\begin{enumerate}
\def\labelenumi{\arabic{enumi})}
\tightlist
\item
  providing an operator algebra framework with domain-specific operators
  for astrophysics and CMB data analysis, exposed through an intuitive,
  math-like interface,
\item
  offering a modular architecture that facilitates experimentation with
  realistic instrument models and complex noise systematics,
\item
  supporting the exploration of novel map-making techniques, including
  integration with \texttt{JAX}-based probabilistic programming tools
  (\citeproc{ref-blackjax}{Cabezas et al., 2024};
  \citeproc{ref-numpyro}{Phan et al., 2019}), which unlocks advanced
  statistical inference methods such as Bayesian hierarchical modeling
  in high-dimensional spaces,
\item
  enabling integration with production pipelines for terabyte-scale
  datasets through GPU acceleration, and supporting hybrid approaches
  that combine neural networks with linear operators for
  simulation-based inference
  (\citeproc{ref-BoeltsDeistler_sbi_2025}{Boelts et al., 2025}).
\end{enumerate}

\section{State of the Field}\label{state-of-the-field}

Most software packages developed for astrophysics and CMB data analysis
target a specific data set, or are limited to a particular step of the
data analysis pipeline, or provide a highly-integrated, end-to-end
framework. Some provide functionality for either data modelling or
analysis only. \texttt{Furax} distinguishes itself from these by
providing a flexible, general, unified, differentiable operator
framework that integrates low-level \texttt{JAX}-compatible libraries
(such as \texttt{jax-healpy} (\citeproc{ref-jax-healpy2024}{Chanial et
al., 2024}) and \texttt{s2fft} (\citeproc{ref-s2fft2024}{Price \&
McEwen, 2024})) as well as interfaces to some other existing tools, such
as \texttt{PySM} (\citeproc{ref-pysm3}{Thorne et al., 2017}) used to
generate realistic multi-component sky simulations.

Few experiment-agnostic CMB data analysis frameworks exist: -
\texttt{TOAST} (\citeproc{ref-toast2021}{Kisner \& Keskitalo, 2021})
provides a comprehensive MPI-parallel modular framework used in
production pipelines for experiments like \emph{Planck} and the Simons
Observatory, but its C++ core does not fully support differentiability
or GPU acceleration, although this has been explored
(\citeproc{ref-demeure2023}{Demeure et al., 2023}). - \texttt{Commander}
(\citeproc{ref-galloway2023beyondplanck}{Galloway et al., 2023}) is a
complementary end-to-end Bayesian framework designed to infer
astrophysical components and cosmological parameters from CMB data, from
maps (and in some implementations timelines) to cosmology, but it
currently lacks GPU support. - At a lower level, \texttt{PyOperators}
(\citeproc{ref-chanial2012pyoperators}{Chanial \& Barbey, 2012})
provides an operator algebra and is used by the QUBIC data analysis
pipeline, and can be seen as a CPU-only precursor to Furax. Similarly,
\texttt{lineax} (\citeproc{ref-kidger2024lineax}{Kidger, 2024}) offers a
\texttt{JAX}-compatible operator algebra but lacks domain-specific
operators and relies on a third-party library for its base operator
class.

Some existing software packages particularly relevant for this work
include: - \texttt{MIDAPACK} (\citeproc{ref-mappraiser2022}{El
Bouhargani et al., 2022}) is a low and middle layer, massively parallel
(MPI) C library for CMB inverse problems, featuring \texttt{MAPPRAISER},
a flexible map-making code, which builds on it - \texttt{FGBuster}
(\citeproc{ref-rizzieri2025}{Rizzieri et al., 2025};
\citeproc{ref-fgbuster2022}{Stompor \& others, 2022}) implements
parametric component separation but relies on simplified noise models

\section{Software Design}\label{software-design}

\texttt{Furax}'s architecture centers on composable linear operators,
which are implemented as Python dataclasses registered as \texttt{JAX}
pytrees. Operators are combined using standard mathematical notation:

\begin{Shaded}
\begin{Highlighting}[]
\NormalTok{H }\OperatorTok{=}\NormalTok{ detector\_response }\OperatorTok{@}\NormalTok{ band\_pass }\OperatorTok{@}\NormalTok{ hwp }\OperatorTok{@}\NormalTok{ rotation }\OperatorTok{@}\NormalTok{ pointing }\OperatorTok{@}\NormalTok{ mixing\_matrix}
\NormalTok{N }\OperatorTok{=}\NormalTok{ HomothetyOperator(σ}\OperatorTok{**}\DecValTok{2}\NormalTok{, in\_structure}\OperatorTok{=}\NormalTok{H.out\_structure)  }\CommentTok{\# Noise covariance}
\NormalTok{m }\OperatorTok{=}\NormalTok{ \{}\StringTok{\textquotesingle{}cmb\textquotesingle{}}\NormalTok{: jnp.random(…), }\StringTok{\textquotesingle{}dust\textquotesingle{}}\NormalTok{: …, }\StringTok{\textquotesingle{}atmosphere\textquotesingle{}}\NormalTok{: …, …\}  }\CommentTok{\# Sky components}
\NormalTok{A }\OperatorTok{=}\NormalTok{ (H.T }\OperatorTok{@}\NormalTok{ N.I }\OperatorTok{@}\NormalTok{ H).I }\OperatorTok{@}\NormalTok{ H.T }\OperatorTok{@}\NormalTok{ N.I  }\CommentTok{\# Sky components\textquotesingle{} maximum{-}likelihood estimator}
\NormalTok{d }\OperatorTok{=}\NormalTok{ H(m) }\OperatorTok{+}\NormalTok{ noise                   }\CommentTok{\# noisy TOD}
\NormalTok{solution }\OperatorTok{=}\NormalTok{ A(d)                    }\CommentTok{\# Inverse via solvers}
\end{Highlighting}
\end{Shaded}

\textbf{Operator algebra.} The base class
\texttt{AbstractLinearOperator} provides a default implementation for
standard linear algebra operations that enable intuitive composition and
manipulation of operators:

\begin{longtable}[]{@{}
  >{\raggedright\arraybackslash}p{(\linewidth - 2\tabcolsep) * \real{0.2667}}
  >{\raggedright\arraybackslash}p{(\linewidth - 2\tabcolsep) * \real{0.7333}}@{}}
\caption{Supported operator operations in
\texttt{Furax}.}\tabularnewline
\toprule\noalign{}
\begin{minipage}[b]{\linewidth}\raggedright
Operation
\end{minipage} & \begin{minipage}[b]{\linewidth}\raggedright
Syntax
\end{minipage} \\
\midrule\noalign{}
\endfirsthead
\toprule\noalign{}
\begin{minipage}[b]{\linewidth}\raggedright
Operation
\end{minipage} & \begin{minipage}[b]{\linewidth}\raggedright
Syntax
\end{minipage} \\
\midrule\noalign{}
\endhead
\bottomrule\noalign{}
\endlastfoot
Addition & \texttt{A\ +\ B} \\
Composition & \texttt{A\ @\ B} \\
Multiplication by scalar & \texttt{k\ *\ A} \\
Transpose & \texttt{A.T} \\
Inverse & \texttt{A.I} or \texttt{A.I(solver=…,\ preconditioner=…)} \\
Block Assembly & \texttt{BlockColumnOperator},
\texttt{BlockDiagonalOperator}, \texttt{BlockRowOperator} \\
\end{longtable}

\textbf{Generic operators.} \texttt{Furax} provides a comprehensive
suite of generic operators for common mathematical operations:

\begin{longtable}[]{@{}
  >{\raggedright\arraybackslash}p{(\linewidth - 2\tabcolsep) * \real{0.3523}}
  >{\raggedright\arraybackslash}p{(\linewidth - 2\tabcolsep) * \real{0.6477}}@{}}
\caption{Generic operators available in \texttt{Furax}.}\tabularnewline
\toprule\noalign{}
\begin{minipage}[b]{\linewidth}\raggedright
Operator
\end{minipage} & \begin{minipage}[b]{\linewidth}\raggedright
Description
\end{minipage} \\
\midrule\noalign{}
\endfirsthead
\toprule\noalign{}
\begin{minipage}[b]{\linewidth}\raggedright
Operator
\end{minipage} & \begin{minipage}[b]{\linewidth}\raggedright
Description
\end{minipage} \\
\midrule\noalign{}
\endhead
\bottomrule\noalign{}
\endlastfoot
\texttt{IdentityOperator} & Returns the input unchanged \\
\texttt{HomothetyOperator} & Multiplication by a scalar \\
\texttt{DiagonalOperator} & Element-wise multiplication \\
\texttt{BroadcastDiagonalOperator} & Non-square operator for
broadcasting \\
\texttt{TensorOperator} & For dense matrix operations \\
\texttt{TreeOperator} & For generalized matrix operations \\
\texttt{SumOperator} & Sum along axes \\
\texttt{IndexOperator} & Can be used for projecting skies onto
time-ordered series \\
\texttt{MaskOperator} & Bit-encoded 0- or 1-valued mask \\
\texttt{MoveAxisOperator} & Manipulate axes of input pytrees \\
\texttt{ReshapeOperator} & Reshape input pytrees \\
\texttt{RavelOperator} & Flatten input pytrees \\
\texttt{FFTOperator} & Fast Fourier transform \\
\texttt{SymmetricBandToeplitzOperator} & Methods: direct convolution,
FFT, overlap and save \\
\texttt{BlockRowOperator} & For horizontal stacking \\
\texttt{BlockDiagonalOperator} & For independent parallel
computations \\
\texttt{BlockColumnOperator} & For vertical stacking \\
\end{longtable}

The block operators provided by the framework enable efficient
structuring of complex multi-observation or multi-component systems.
\texttt{SymmetricBandToeplitzOperator} provides efficient convolution
operations using the overlap-save method. This operator is central to
correlated noise modeling and gap-filling procedures based on
constrained Gaussian realizations (\citeproc{ref-stompor2002}{Stompor et
al., 2002}).

\textbf{Domain-specific operators.} For CMB data analysis,
\texttt{Furax} includes specialized operators tailored to instrument
modeling and astrophysical components:

\begin{longtable}[]{@{}ll@{}}
\caption{Domain-specific operators for astrophysics or CMB data
analysis.}\tabularnewline
\toprule\noalign{}
Operator & Description \\
\midrule\noalign{}
\endfirsthead
\toprule\noalign{}
Operator & Description \\
\midrule\noalign{}
\endhead
\bottomrule\noalign{}
\endlastfoot
\texttt{QURotationOperator} & Stokes QU rotation \\
\texttt{HWPOperator} & Ideal half-wave plate \\
\texttt{LinearPolarizerOperator} & Ideal linear polarizer \\
\texttt{CMBOperator} & Parametrized CMB SED \\
\texttt{DustOperator} & Parametrized dust SED \\
\texttt{SynchrotronOperator} & Parametrized synchrotron SED \\
\texttt{PointingOperator} & On-the-fly projection matrix \\
\texttt{MapSpaceBeamOperator} & Sparse Beam operator \\
\texttt{TemplateOperator} & For template map-making \\
\end{longtable}

For instance, \texttt{HWPOperator} is used for half-wave plate modeling,
\texttt{LinearPolarizerOperator} for polarization extraction and
\texttt{QURotationOperator} for polarization angle rotations. The
spectral operators (\texttt{CMBOperator}, \texttt{DustOperator},
\texttt{SynchrotronOperator}) enable frequency-dependent component
separation with support for spatially varying spectral indices.

\textbf{Algebraic reductions.} \texttt{Furax} implements operator
simplification through a rule-based system, complementing XLA's
low-level optimizations (\citeproc{ref-xla2017}{Google, 2017}). For
example, consecutive QU rotations combine their angles, and compositions
involving block operators such as \(P^\top N^{-1} P\) are decomposed
into \(\sum_i P_i^\top N_i^{-1} P_i\), exploiting block structure to
reduce computational cost. The system also handles algebraic identities
such as the commutation rule for ideal half-wave plates:
\(R(\theta) \circ \text{HWP} = \text{HWP} \circ R(-\theta)\).

\textbf{Stokes parameter types.} \texttt{Furax} represents polarization
through dedicated \texttt{JAX} pytrees: \texttt{StokesI},
\texttt{StokesQU}, \texttt{StokesIQU}, and \texttt{StokesIQUV}. These
types support arithmetic operations, broadcasting, and seamless
integration with \texttt{JAX} transformations.

\textbf{Landscape types.} They specify how input signals are discretized
on the sky (e.g., HEALPix spherical pixelization) and provide the
mapping to world coordinates.

\textbf{Map-making classes.} \texttt{Furax} features several map-making
algorithms, including the binned estimator, maximum-likelihood, and
template deprojection methods. \texttt{PointingOperator} implements the
sky-to-TOD projection and rotation into the instrument frame, expanding
the time-dependent pointing information on the fly to save memory.
\texttt{MultiObservationMapMaker} provides simultaneous reconstruction
across multiple observations without recompiling the compute kernels.
\texttt{BJPreconditioner} implements the block Jacobi preconditioner to
accelerate convergence of the numerical inverse.

\section{Research Impact Statement}\label{research-impact-statement}

\texttt{Furax} was developed within the
\href{https://scipol.in2p3.fr}{\textsc{SciPol} project} to enable
GPU-accelerated and gradient-based optimization in CMB data analysis
pipelines. The framework's differentiability opens new possibilities for
neural network integration and end-to-end optimization of map-making and
component separation. The modular design supports rapid prototyping of
analysis methods while maintaining compatibility with production
pipelines through \texttt{TOAST} integration. \texttt{Furax} provides
essential infrastructure for developing next-generation analysis
techniques for e.g., the Simons Observatory, the South Pole Observatory,
QUBIC and \emph{LiteBIRD}.

\section{AI Usage Disclosure}\label{ai-usage-disclosure}

AI-assisted tools were used for code documentation and manuscript
preparation. All AI-generated content was verified by the authors.

\section{Acknowledgements}\label{acknowledgements}

\texttt{Furax} draws inspiration from \texttt{PyOperators}
(\citeproc{ref-chanial2012pyoperators}{Chanial \& Barbey, 2012}) and
\texttt{lineax} (\citeproc{ref-kidger2024lineax}{Kidger, 2024}).

This work was supported by the European Research Council (ERC) under the
European Union's Horizon 2020 research and innovation programme (Grant
Agreement No.\textasciitilde101044073, PI: Josquin Errard).

Computing resources were provided by GENCI at IDRIS (Jean Zay
supercomputer) under allocations 2024-AD010414161R2 and
2025-A0190416919.

This work has also received funding by the European Union's Horizon 2020
research and innovation program under grant agreement no. 101007633
CMB-Inflate.

MM is funded by the European Union (ERC, RELiCS, project number
101116027). Views and opinions expressed are however those of the
authors only and do not necessarily reflect those of the European Union
or the European Research Council Executive Agency. Neither the European
Union nor the granting authority can be held responsible for them.

\section*{References}\label{references}
\addcontentsline{toc}{section}{References}

\protect\phantomsection\label{refs}

\end{document}